\documentclass[12pt]{article}
\newlength{\pubnumber} 

\catcode`\@=11
\@addtoreset{equation}{section}

\def\section{\@startsection{section}{1}{\z@}{3.5ex plus 1ex minus .2ex}
 {2.3ex plus .2ex}{\large\bf}}
\def\subsection{\@startsection{subsection}{2}{\z@}{2.3ex plus .2ex}
 {2.3ex plus .2ex}{\bf}}

 \usepackage{graphicx}
\begin{document}

\begin{titlepage}
\samepage{
\setcounter{page}{0}
\rightline{April 2007}
\vfill
\begin{center}
    {\Large \bf Supersymmetry versus Gauge Symmetry on the Heterotic Landscape\\}
\vfill
\vspace{.10in}
   {\large
      Keith R. Dienes$^1$\footnote{
     E-mail address:  dienes@physics.arizona.edu},
    Michael Lennek$^1$\footnote{E-mail address:  mlennek@physics.arizona.edu}, 
     David S\'en\'echal$^2$\footnote{E-mail address:  david.senechal@usherbrooke.ca},\\ 
  Vaibhav Wasnik$^1$\footnote{E-mail address:  wasnik@physics.arizona.edu} \\}
\vspace{.13in}
 {\it  $^1\,$Department of Physics, University of Arizona, Tucson, AZ  85721  USA\\}
 {\it  $^2\,$D\'epartement de Physique, Universit\'e de Sherbrooke, Sherbrooke, Qu\'ebec J1K 2R1  Canada\\}
\end{center}
\vfill
\begin{abstract}
  {\rm  One of the goals of the landscape program in string theory is to 
        extract information about the space of string vacua
        in the form of statistical correlations between phenomenological features
        that are otherwise uncorrelated in field theory.
        Such correlations would thus represent predictions of string theory that
        hold independently of a vacuum-selection principle.
        In this paper, we study statistical correlations between
        two features which are likely to be central to any potential
        description of nature at high energy scales:  gauge symmetries 
        and spacetime supersymmetry.
        We analyze correlations between these two kinds of symmetry within the 
        context of perturbative heterotic string vacua, and 
        find a number of striking features.
        We find, for example, 
        that the degree of spacetime supersymmetry
        is strongly correlated with the probabilities of realizing
        certain gauge groups, with unbroken supersymmetry at the string scale
        tending to favor gauge-group factors with larger rank.
        We also find 
        that nearly half of the heterotic landscape is 
        non-supersymmetric and yet tachyon-free at tree level;  indeed, less than a quarter
        of the tree-level heterotic landscape exhibits any supersymmetry at all at the string scale.
  }
\end{abstract}
\vfill
\smallskip}
\end{titlepage}

\setcounter{footnote}{0}

\def\beq{\begin{equation}}
\def\eeq{\end{equation}}
\def\beqn{\begin{eqnarray}}
\def\eeqn{\end{eqnarray}}
\def\half{{\textstyle{1\over 2}}}

\def\calO{{\cal O}}
\def\calE{{\cal E}}
\def\calT{{\cal T}}
\def\calM{{\cal M}}
\def\calF{{\cal F}}
\def\calY{{\cal Y}}
\def\calV{{\cal V}}
\def\calN{{\cal N}}
\def\ibar{{\overline{\i}}}
\def\qbar{{\overline{q}}}
\def\mm{{\tilde m}}
\def\ahat{{\hat a}}
\def\nn{{\tilde n}}
\def\rep#1{{\bf {#1}}}
\def\ie{{\it i.e.}\/}
\def\eg{{\it e.g.}\/}

\def\Str{{{\rm Str}\,}}
\def\bone{{\bf 1}}

\def\thetai{{\vartheta_i}}
\def\thetaj{{\vartheta_j}}
\def\thetak{{\vartheta_k}}
\def\thetaibar{\overline{\vartheta_i}}
\def\thetajbar{\overline{\vartheta_j}}
\def\thetakbar{\overline{\vartheta_k}}
\def\etainv{{\overline{\eta}}}

\def\modinvmeasure{{  {{{\rm d}^2\tau}\over{\tautwo^2} }}}
\def\qbar{{  \overline{q} }}
\def\ahat{{ \hat a }}

\newcommand{\newc}{\newcommand}
\newc{\gsim}{\lower.7ex\hbox{$\;\stackrel{\textstyle>}{\sim}\;$}}
\newc{\lsim}{\lower.7ex\hbox{$\;\stackrel{\textstyle<}{\sim}\;$}}

\hyphenation{su-per-sym-met-ric non-su-per-sym-met-ric}
\hyphenation{space-time-super-sym-met-ric}
\hyphenation{mod-u-lar mod-u-lar--in-var-i-ant}


\def\inbar{\,\vrule height1.5ex width.4pt depth0pt}

\def\IC{\relax\hbox{$\inbar\kern-.3em{\rm C}$}}
\def\IQ{\relax\hbox{$\inbar\kern-.3em{\rm Q}$}}
\def\IR{\relax{\rm I\kern-.18em R}}
 \font\cmss=cmss10 \font\cmsss=cmss10 at 7pt
\def\IZ{\relax\ifmmode\mathchoice
 {\hbox{\cmss Z\kern-.4em Z}}{\hbox{\cmss Z\kern-.4em Z}}
 {\lower.9pt\hbox{\cmsss Z\kern-.4em Z}}
 {\lower1.2pt\hbox{\cmsss Z\kern-.4em Z}}\else{\cmss Z\kern-.4em Z}\fi}

\long\def\@caption#1[#2]#3{\par\addcontentsline{\csname
  ext@#1\endcsname}{#1}{\protect\numberline{\csname
  the#1\endcsname}{\ignorespaces #2}}\begingroup
    \small
    \@parboxrestore
    \@makecaption{\csname fnum@#1\endcsname}{\ignorespaces #3}\par
  \endgroup}
\catcode`@=12

\input epsf

\section{Introduction}
\setcounter{footnote}{0}
\label{intro}

Recent developments in string theory suggest that there exists
a huge ``landscape'' of self-consistent string vacua~\cite{landscape}.  
The existence of this landscape is of critical  
importance for string phenomenology
since the specific low-energy phenomenology that
can be expected to emerge from string theory depends
critically on the particular choice of vacuum state.
Detailed quantities such as particle masses and mixings,
and even more general quantities and structures such as the
choice of gauge group, number of chiral particle generations,
and the magnitude of the supersymmetry-breaking scale,
can be expected to vary significantly from one vacuum solution to the next.
Thus, in the absence of some sort of vacuum-selection principle,
it is natural to determine whether there might exist generic 
string-derived {\it statistical correlations}\/
between different phenomenological features that would otherwise be uncorrelated
in field theory~\cite{abstract}.
In this way, one can still hope to extract phenomenological predictions
from string theory.

To date, there has been considerable work in this
direction~\cite{abstract,abstract2,susybreakingabstract,direct,direct2,direct3,dienes,
Measure,Raby,Faraggi,fieldtheory};
for recent reviews, see Ref.~\cite{reviews}.
Collectively, this work addresses questions
ranging from the formal (such as the finiteness of the number of string vacua and the
methods by which they may be efficiently scanned and classified) 
to the phenomenological (such as the
value of the cosmological constant, the scale of supersymmetry breaking, 
and the statistical prevalence of the
Standard Model gauge group and three chiral generations).

In this paper, we shall undertake a statistical study of the correlations between
two phenomenological
features which are likely to be central to any description of nature at high energy scales:
spacetime supersymmetry and gauge symmetry.
Indeed, over the past twenty years, a large amount of theoretical effort has been 
devoted to studying string models with $\calN{=} 1$ spacetime 
supersymmetry.  However, it is important to understand the implications of 
choosing $\calN{=} 1$ supersymmetry over other classes of string models (such as 
models with $\calN{=} 2$ or $\calN{=}4$ supersymmetry, or even non-supersymmetric
string models)  within the context of the landscape.  
Moreover, since $\calN{=}1$ supersymmetry plays a huge role in current theoretical efforts
to extend the Standard Model,  
we shall also be interested in understanding the statistical prevalence of
spacetime supersymmetry across the landscape and the degree to which the
presence or absence of supersymmetry
affects other phenomenological features such as the choice of gauge group
and the resulting particle spectrum.

In this paper, we shall investigate such questions within the context of the
heterotic string landscape.
There are several reasons why we shall focus on the heterotic landscape.
First, heterotic strings are of tremendous phenomenological interest in their own right;
indeed, these strings the framework in which most of the original work in string phenomenology
was performed in the late 1980's and early 1990's.
Second, heterotic strings have internal constructions and self-consistency constraints
which are, in many ways, more constrained than those of their Type~I (open) counterparts.
Thus, they are likely to exhibit phenomenological correlations which differ from those
that might be observed on the landscape of, say, intersecting D-brane models or
Type~I flux vacua.
Finally, in many cases these perturbative supersymmetric heterotic strings are dual
to other strings (\eg, Type~I orientifold models) whose statistical
properties are also being analyzed in the literature.  Thus, analysis of the perturbative
heterotic landscape, both supersymmetric and non-supersymmetric, might 
eventually enable {\it statistical}\/ tests of duality symmetries across
the entire string landscape.

The first statistical study of the heterotic landscape
appeared in Ref.~\cite{dienes}. 
This study, which focused exclusively on the statistical properties 
of non-supersymmetric ($\calN{=}0$) tachyon-free heterotic string vacua,
was based on a relatively small
data set of four-dimensional heterotic string models~\cite{PRL}
which were randomly generated using software originally developed
in Ref.~\cite{Senechal}.
Since then, there have been several additional statistical
examinations of certain classes of $\calN{=}1$ supersymmetric heterotic strings~\cite{Raby,Faraggi}. 
Together, such studies can therefore be viewed as  providing heterotic analogues
of the Type~I statistical studies reported in Refs.~\cite{direct,direct2,direct3}.

Although the study we shall undertake here is similar in spirit to that of Ref.~\cite{dienes}, 
there are several important differences which must be highlighted.
First, as discussed above, we shall be focusing here on the effects of spacetime
supersymmetry.  Thus, we shall be examining models with all levels of spacetime
supersymmetry ($\calN{=}0,1,2,4$), not just non-supersymmetric models, and 
examining how the level of spacetime supersymmetry
correlates with gauge symmetry.
Second, the current study is based on a much larger data set consisting of 
approximately $10^7$ heterotic string models which was newly generated for this purpose
using an update of the software originally developed in Ref.~\cite{Senechal}.  
This data set is thus approximately
two orders of magnitude larger than that used for Ref.~\cite{dienes}, and represents literally the largest
set of distinct heterotic string models ever constructed.
Indeed, for reasons we shall discuss in Sect.~\ref{Analysis}, we believe
that data sets of this approximate size are probably among the largest that can
be generated using current computer technology.

But perhaps most importantly, because our heterotic-string data set was 
newly generated for the purpose of this study, 
we are able to quote results that take into account
certain subtleties concerning so-called ``floating correlations''.
As discussed in Ref.~\cite{Measure},
the problem of floating correlations 
is endemic to investigations of this type,
and reflects the fact that not all physically distinct
string models are equally likely to be sampled in any random
search through the landscape.  This  thereby causes statistical correlations
to ``float'' as a function of sample size.
In Ref.~\cite{Measure}, 
   several methods were developed that can be used
   to overcome this problem, and it was shown 
    through explicit examples that these methods
    allow one to extract correlations and statistical distributions
    which are not only stable as a function of sample size,
         but which also differ significantly from those which
    would have been na\"\i vely apparent from a direct counting of generated models.
We shall therefore employ these techniques in the current paper,
extracting each of our statistical results in such a way that
they represent stable correlations across the entire 
heterotic landscape we are examining.

As with most large-scale statistical studies of this type,
there are several limitations which must be borne in mind.
First, our sample size is relatively small, consisting of
only $\sim 10^7$ distinct models.   However, although this number is
miniscule compared with the numbers of string models that are currently
quoted in most landscape discussions, we believe that the statistical results
we shall obtain are stable as a function of sample size and would not change 
significantly as more models are added to the data sample.
We shall discuss this feature in more detail in Sect.~\ref{Analysis}.
Indeed, as mentioned above, data samples of the
current size are likely to be the largest possible given current computer technology.

Second, the analysis in this paper shall be limited to correlations between only
two phenomenological properties of these models:  their low-energy gauge groups, and 
their levels of supersymmetry.  More detailed examinations of the particle spectra
of these models will be presented in Ref.~\cite{toappear}.

Finally,
the models we shall be discussing are stable only at tree level.
For example, the models with spacetime supersymmetry continue to have flat 
directions which have not been lifted.  Even worse, the
non-supersymmetric models (even though tachyon-free) will
generally have non-zero dilaton tadpoles and thus
are not stable beyond tree level.
Despite these facts, each of the string models we shall be studying
represents a valid string solution at tree level, satisfying 
all of the necessary string self-consistency constraints.
These include
the requirements of worldsheet conformal/superconformal invariance,
modular-invariant one-loop and multi-loop amplitudes,
proper spacetime spin-statistics relations,
and physically self-consistent layers of sequential GSO projections and orbifold twists.
Thus, although such models may not represent
the sorts of truly stable vacua that we would ideally like to be studying,
it is reasonable to hope that any statistical correlations 
we uncover are likely to hold even after vacuum stabilization.
Indeed, 
since no stable perturbative non-supersymmetric heterotic strings have
yet been constructed, this sort of analysis is currently the state of the art
for large-scale statistical studies of this type,
and mirrors the situation on the Type~I side, where state-of-the-art
statistical analyses~\cite{direct,direct2,direct3}
have also focused on models which are only stable at tree level.
Eventually, once the heterotic model-building technology 
develops further and truly stable vacua can be analyzed,
it will be interesting to compare those results 
with these in order to ascertain the degree to which vacuum stabilization
might affect these other phenomenological properties.

This paper is organized as follows.
In Sect.~\ref{Models}, we describe the class of 
models that we shall be examining in this paper.  In Sect.~\ref{Analysis},
we summarize our method of analysis which enables us to overcome
the problem of floating correlations in order to extract statistically
meaningful correlations.
In Sect.~\ref{General}, we present our results concerning
the prevalence of spacetime supersymmetry across the heterotic landscape,
and in 
Sect.~\ref{Specific} we present our results concerning correlations
between spacetime supersymmetry and gauge groups. 
Finally, our conclusions are presented in Sect.~\ref{Conclusions}.

\section{The models}
\setcounter{footnote}{0}
\label{Models}

The models we shall be examining in this paper are similar to those studied
in Ref.~\cite{dienes}.
Specifically, each of the vacua we shall be examining in this paper
represents a weakly coupled critical heterotic string compactified to
four large (flat) spacetime dimensions.  
In general, such a string may be described in terms of its left- and right-moving 
worldsheet conformal field theories (CFT's).
For a string in four dimensions, these must have central charges $(c_R,c_L)=(9,22)$ in 
order to enforce worldsheet conformal anomaly cancellation, and must exhibit
conformal invariance for the left-movers and superconformal invariance for the
right-movers.
While any such CFT's may be 
considered, in this paper we shall focus on those string models 
for which these internal worldsheet CFT's 
may be taken to consist of tensor products of free, 
non-interacting, complex (chiral) bosonic or fermionic fields. 

As discussed in Ref.~\cite{dienes}, 
this is a huge class of models which has been discussed and analyzed in many different ways in the string 
literature.  On the one hand, taking these worldsheet fields as fermionic leads to the so-called   
``free-fermionic'' 
construction~\cite{KLT} which will be our primary tool throughout this paper.  In the language of 
this construction, different models are achieved by varying (or ``twisting'') the boundary conditions of these 
fermions around the two non-contractible loops of the worldsheet torus while simultaneously varying the 
phases according to which the contributions of each such spin-structure sector are summed in producing 
the one-loop partition function.  However, alternative but equivalent languages for constructing such models 
exist.  For example, we may bosonize these worldsheet fermions and construct ``Narain'' 
models~\cite{Narain,Lerche} 
in which the resulting complex worldsheet bosons are compactified on internal lattices of 
appropriate dimensionality with appropriate self-duality properties.  
Furthermore, many of these models 
have additional geometric realizations as orbifold compactifications with appropriately 
chosen Wilson lines;  in 
general, the process of orbifolding is quite complicated in these models, involving many sequential layers of 
projections and twists.  All of these constructions generally overlap to a large degree, and all are 
capable of producing models in which the corresponding gauge groups and particle contents are quite 
intricate.  Nevertheless, in all cases, we must ensure that all required self-consistency constraints are 
satisfied.  These include modular invariance, physically sensible GSO projections, proper spin-statistics 
identifications, and so forth.  Thus, each of these vacua represents a fully self-consistent string solution at 
tree level.

In order to efficiently survey the space of such four-dimensional string-theoretic vacua, we implemented a 
computer search based on the free-fermionic spin-structure construction~\cite{KLT}.  
Details of this study are similar to those of the earlier study described in Ref.~\cite{dienes}, and utilize
an updated version of the model-generating software that was originally written
for Ref.~\cite{Senechal}. 
In our analysis, we restricted our attention to those models 
for which our real worldsheet fermions can always be 
uniformly paired to form complex fermions, and therefore it was 
possible to specify the boundary conditions (or spin-structures) of these real fermions in terms of the 
complex fermions directly.  We also restricted our attention to cases in which the worldsheet fermions 
exhibited either antiperiodic (Neveu-Schwarz) or periodic (Ramond) boundary conditions around
the non-contractible loops of the torus.  
Of course, in 
order to build a self-consistent string model in this framework, these boundary conditions must satisfy tight 
constraints.  These constraints are necessary in order to ensure that the one-loop partition function is 
modular invariant and that the resulting Fock space of states can be interpreted as arising 
through a physically 
sensible projection from the space of all worldsheet states onto the subspace of physical states with proper 
spacetime spin-statistics.  Thus, within a given string model, it is necessary to sum over appropriate sets of 
untwisted and twisted sectors with different boundary conditions and projection phases.

Our statistical analysis consisted of an examination of over $10^7$ distinct vacua in this class.
Essentially, each set of fermion boundary conditions and GSO projection phases
was chosen randomly in each sector, subject 
only to the required self-consistency constraints.  However, in our 
statistical sampling, we placed essentially no limits on the complexity of the orbifold 
twisting (\ie, in the free-fermionic language, we allowed as many as sixteen 
linearly independent basis vectors).  Thus, our statistical analysis 
included models of arbitrary intricacy and sophistication.  We also made use of techniques developed 
specifically for analyzing string models generated in random searches, allowing for the mitigation of many of 
the effects of bias which are endemic to studies of this sort.

As part of our study,
we generated string models with all degrees of spacetime supersymmetry 
($\calN{=}0,1,2,4$) that can arise in four dimensions.
For $\calN{=} 0$ models, we further demanded that supersymmetry be 
broken without introducing tachyons.  Thus, the $\calN{=}0$ vacua are all non-supersymmetric but tachyon-free, 
and can be considered as four-dimensional analogues of the ten-dimensional $SO(16)\times SO(16)$ heterotic 
string~\cite{SOsixteen} which is also non-supersymmetric but tachyon-free.  However, other than 
this, we  
placed no requirements on other possible  phenomenological properties of these vacua such as their 
possible gauge groups, numbers of chiral generations, or other aspects of the particle content.  We did, 
however, require that our string construction  begin with a supersymmetric theory in which the supersymmetry
may or may not be broken by subsequent orbifold twists.  (In the language of the 
free-fermionic construction, this is tantamount to demanding that our fermionic boundary conditions include 
a superpartner sector, typically denoted ${\bf W}_1$ or ${\bf V}_1$.)  This is to be distinguished from a 
potentially more general class of models in which supersymmetry does not appear at any stage of the 
construction.  This is merely a technical detail in our construction, and we do not believe that this ultimately 
affects our results.

As with any string-construction method, the free-fermionic formalism contains numerous
redundancies in which different choices of worldsheet fermion boundary conditions
and/or GSO phases lead to identical string models in spacetime.  
Indeed, a given unique string model can have many different representations
in terms of worldsheet constructions.
For this reason, we judged string vacua to 
be distinct based on their spacetime characteristics --- 
 {\it i.e.}\/, their low-energy gauge groups and 
massless particle content.

\begin{table}[htb]
\begin{center}
\begin{tabular}{||c|r||}
\hline
\hline
SUSY class & $\#$ distinct models \\
\hline 
$\calN {=} 0$ (tachyon-free)&   4$\,$946$\,$388  \\
$\calN {=} 1$ &    3$\,$772$\,$679     \\
$\calN {=} 2$ &   492$\,$790     \\
$\calN {=} 4$ &  1106              \\
\hline
Total: & 9$\,$212$\,$963  \\
\hline
\hline
\end{tabular}
\end{center}
\caption{
The data set of perturbative heterotic strings analyzed in this paper.
For each level of supersymmetry allowed in four dimensions,
we list the number of corresponding distinct models generated.
As discussed in the text,
models are judged to be distinct 
based on their spacetime properties (\eg, gauge groups and particle content).
All non-supersymmetric models listed here are tachyon-free and thus are four-dimensional 
analogs of the $SO(16) \times SO(16)$ string model in ten dimensions.}  
\label{firsttable}
\end{table}

Given this, our ultimate data set of heterotic strings
is as described in Table~\ref{firsttable}.
Note that all non-supersymmetric models listed in Table~\ref{firsttable}
are tachyon-free, and thus are stable at tree level.
We should mention that while generating these models,
we also generated over a million distinct non-supersymmetric
tachyonic vacua which are not even stable at tree level.
We therefore did not include their properties in our analysis,
and recorded their existence only as a way of gauging the overall degree
to which the tree-level heterotic string landscape is tachyon-free.  
Also note that as the level of supersymmetry increases, the number of distinct
models in our sample set decreases. 
This reflects the fact that relatively fewer of these models 
exist, so they become more and more difficult to generate.
This will be discussed further in Sects.~3 and 4.

Of course, the free-fermionic construction realizes only 
certain points in the full model space of self-consistent heterotic string models.  For example, since each 
worldsheet fermion is nothing but a worldsheet boson compactified at a specific radius, a larger (infinite) 
class of models can immediately be realized through a bosonic formulation by varying these radii away from 
their free-fermionic values.  However, this larger class of models has predominantly only abelian 
gauge groups and 
rather limited particle representations.  Indeed, the free-fermionic points typically represent 
precisely those points at which additional (non-Cartan) gauge-boson states become massless, thereby 
enhancing the gauge symmetries to become non-abelian.  Thus, the free-fermionic construction naturally 
leads to precisely the set of models which are likely to be of direct phenomenological relevance.

We should note that 
it is also possible to go beyond the class of free-field string models altogether, and consider models built 
from more complicated worldsheet CFT's (\eg, Gepner models).  One could even go beyond the model 
space of critical string theories, and consider non-critical strings and/or strings with non-trivial background 
fields.  Likewise, we may consider heterotic strings beyond the usual perturbative limit.  However, although 
such models may well give rise to phenomenologies very different from those that emerge in free-field 
constructions, their spectra are typically very difficult to analyze and are thus not amenable to an automated 
statistical investigation.

\section{Method of analysis}
\label{Analysis}
\setcounter{footnote}{0}

Each string model-construction technique provides a mapping between
a space of internal parameters 
and a corresponding physical string model in spacetime.  
In the case of closed strings, for example,
such internal parameters might include
compactification moduli, boundary-condition phases, Wilson-line coefficients,
or topological quantities specifying Calabi-Yau manifolds;
in the case of open strings, by contrast, they might include D-brane dimensionalities and charges, 
wrapping numbers or intersection angles,
fluxes, and the vevs of moduli fields.
Regardless of the construction technique at hand, however, there is a well-defined
procedure through which one can derive the spectrum and couplings of
the corresponding model in spacetime. 

Given this, one generally conducts a random search through the space of models
by randomly choosing self-consistent values of these internal parameters,
and then deriving the physical properties of the corresponding string models.  
Questions about statistical correlations are then addressed in terms
of the relative abundances of models that emerge with different spacetime characteristics.
Indeed, if $\lbrace \alpha,\beta,\gamma,...\rbrace$ denote these different
spacetime characteristics 
(or different combinations of these characteristics), then we are generally
interested in extracting ratios of population abundances of the form 
$N_\alpha/N_\beta$, where $N_\alpha$ and $N_\beta$ are the numbers of models
which exhibit physical characteristics $\alpha$ and $\beta$ across the landscape
as a whole. 

Clearly, we cannot survey the entire landscape, and thus we are forced to attempt
to extract such ratios with relatively limited information.
In particular, let us assume that our search has consisted of
analyzing $D$ different randomly generated sets of internal parameters,
ultimately yielding a set of different models in spacetime exhibiting varying
physical characteristics.  Let $M_\alpha(D)$ denote the number of
distinct models which are found which exhibit characteristic $\alpha$.
Our natural tendency is then to attempt to associate
\beq
         {N_\alpha\over N_\beta}  ~~{\stackrel{?}{=}}~~ {M_\alpha(D)\over M_\beta(D)}~
\label{naive}
\eeq 
for some sufficiently large value of $D$.
While this relation might not hold exactly for relatively small values of $D$,
the expectation is that we might be able to reach sufficiently large values of $D$ for which
we might hope to extract reasonably accurate predictions for $N_\alpha/N_\beta$.

Unfortunately, as has recently been discussed in Ref.~\cite{Measure},
Eq.~(\ref{naive}) does not generally hold for any reasonable value of $D$ (short
of exploring the full landscape).
Indeed, the violations of this relation are striking, even in situations in
which sizable fractions of the landscape are explored, and will ultimately doom
any attempt at extracting population fractions in this manner.
In the remainder of this section, we shall first explain why 
Eq.~(\ref{naive}) fails.
We shall then summarize the methods which were developed in Ref.~\cite{Measure} for
circumventing these difficulties, and which we will be employing in the
remainder of this paper. 

As stated above,
each string model-construction technique 
provides a mapping between a space of internal parameters and a physical
string model in spacetime.  
However, this mapping is not one-to-one, and there
generally exists a huge redundancy wherein a single physical string model 
in spacetime can have multiple realizations
or representations
in terms of internal parameters.  For this reason, the space of internal parameters
is usually significantly larger than the space of obtainable distinct models.

The failure of this mapping to be one-to-one is critical because
any random statistical study of the string landscape must ultimately take
the form of a random exploration of the space of internal parameters that 
lead to these models.
First, one must randomly choose a self-consistent configuration 
of internal parameters;  only then can one derive and tabulate the spacetime 
properties of the corresponding model.
But then we are faced with the question of determining whether
spacetime models with multiple internal realizations should be weighted more
strongly in our statistical analysis than models with relatively 
few realizations.
In other words, we must decide whether our landscape {\it measure}\/
should be based on internal parameters
(wherein each model is weighted according to its number of internal
realizations)
or based on spacetime properties (wherein each physically distinct model is
weighted equally regardless of the number of its internal realizations).

If we were to base our landscape measure on internal parameters,
then these redundancies would not represent problems;
they would instead become vital ingredients in our numerical
analysis.
However, 
if we are to perform statistics in the space of models
in a physically significant way,
it is easy to see that   
we are forced to count distinct models rather than distinct combinations of 
internal parameters. 
The reason for this is as follows.
In many cases, these redundancies 
arise as the result of worldsheet symmetries (\eg, mirror symmetries),
and even though such symmetries may be difficult to analyze and eliminate
analytically for reasonably complicated models, their associated
redundancies are similar to the redundancies of gauge transformations 
and do not represent new physics.
In other cases, such redundancies are simply 
reflections of the 
failures or limitations of a particular
model-construction technique;  once again, however,
they do not represent new physics, but rather reflect a poor choice
of degrees of freedom for our internal parameters, or a mathematical difficulty
or inability
to properly define their independent domains.
Finally, such redundancies can also emerge because
entirely different model-construction techniques 
can often lead to identical models in spacetime.
Thus, two landscape researchers using different construction formalisms
might independently generate random sets of models which partially overlap, but
once again this does not mean that the models which are common to both sets
should be double-counted when their statistical results are merged.
Indeed, in all of these cases,
redundancies in the mapping between internal parameters and spacetime
properties do not represent differences of physics,
but rather differences in the description of that physics.
We thus must use spacetime characteristics (rather than
the parameters internal to a given string construction) as our means of
counting and distinguishing string models.

Many of these ideas can be illustrated by considering
the $E_8\times E_8$ heterotic string
in ten dimensions.  As is well known, this string model can 
be represented in many ways:
as a $\IZ_2$ orbifold
of the $SO(32)$ supersymmetric string, as a 
$\IZ_2\times \IZ_2$ orbifold of the non-supersymmetric $SO(32)$ heterotic string,
and so forth.  Likewise, this model can be realized
through an orbifold construction,
through a free-fermionic construction, 
through a bosonic lattice construction, and through other constructions
as well.  Yet, there is only a single $E_8\times E_8$ string model
in ten dimensions.
It is therefore necessary to tally distinct 
string models, and not distinct internal formulations,
when performing landscape calculations and interpreting their results.

Unfortunately,
this redundancy inherent in the mapping between internal parameters and
their corresponding string models
implies that in any random exploration of the space of models,
certain string models 
are likely to be sampled
much more frequently than other models.
Thus, one must filter out this effect by keeping a record of each distinct
model that has already been sampled so that each time an additional model is generated
(\ie, each time there is a new ``attempt''),
it can be compared against all previous models and discarded if it is not new.
Although this is a memory-intensive and time-consuming process which ultimately
limits the sizes of the resulting data sets that can be generated using
current automated technology, this filtering can successfully be employed
to eliminate model redundancies.

However, there remains the converse problem:  because some models 
strongly dominate the random search,
others effectively recede and are therefore extremely difficult to reach.  
They therefore do not tend to show up during the early stages of a random
search, and tend to emerge only later in the search process after the dominant
models have been more fully tallied.
Indeed, as the search proceeds into its later stages, 
it is only the models with ``rare'' characteristics
which increasingly tend to be generated, precisely because those models
with ``common'' characteristics
will have already been generated and tabulated.
Thus, the proportion of models with ``rare'' characteristics tends to evolve
rather dramatically as a function of time through the model-generation process. 

This type of bias is essentially unavoidable, 
and has the potential to
seriously distort the values of any numerical correlations that might be extracted from 
a random search through the landscape.
In particular, as discussed in Ref.~\cite{Measure}, this type of bias
generally causes statistical correlations to ``float'' or evolve
as a function of the sample size of models examined.
Moreover, since one can ultimately explore only a limited portion of the landscape,
there is no opportunity to gather statistics at the endpoint of the search
process at which these correlations would have floated to their true values.
This, then, is the problem of floating correlations.

Fortunately, as discussed in Ref.~\cite{Measure}, there are several
statistical methods which can be used in order to overcome this difficulty.
These methods enable one to
extract statistical correlations and distributions which are stable
as a function of sample size and which, with some reasonable assumptions, 
represent the statistical results that would be obtained if the
full space of models could be explored.
We shall now describe the most important of these methods, since we
shall be using this technique throughout the rest of this paper.

In general, a model search proceeds as follows.
One randomly generates a self-consistent set of internal parameters, 
and calculates the properties of the corresponding string model.
One then compares this model against all models which have previously
been generated:  if the model is distinct, it is recorded and saved;
if it is redundant, it is discarded.
One then repeats this process.
Early in the process, most attempts result in new distinct models
because very few models have already been found.
However, as the search proceeds, an increasing fraction of attempts
fail to produce new models.  This rise in the ratio of attempts per new model
indicates that the space of models is becoming more and more explored.
Thus, attempts per model can be used as a measure of how far into the
full space of corresponding models our search has penetrated.

Therefore, if we are 
interested in extracting the ratio
$N_\alpha/N_\beta$ for two physical characteristics
$\alpha$ and $\beta$, 
as discussed above Eq.~(\ref{naive}), 
the solution is {\it not}\/ to extract this ratio
through Eq.~(\ref{naive}) because such a relation assumes
that the spaces of $\alpha$-models and $\beta$-models
are being penetrated at exactly the same rates during the random search
process.  Rather, the solution~\cite{Measure} is to keep a record
not only of the models generated as the search proceeds, 
but also of the cumulative average {\it attempts}\/ per model
that are needed in order to generate these models.  We then extract
the desired ratio $N_\alpha/N_\beta$ through
a relation of the form
\beq
        ~{N_\alpha\over N_\beta}  ~=~ {M_\alpha(d_\alpha)\over M_\beta(d_\beta)} 
             \Bigg |_{  
                  {d_\alpha\over M_\alpha(d_\alpha)} =            
                  {d_\beta\over M_\beta(d_\beta)} } 
\label{kgolden2}
\eeq
where $d_{\alpha}$ and $d_\beta$ respectively represent the numbers of attempts that resulted 
in $\alpha$-models and $\beta$-models, regardless of whether the models in each class
were distinct.
Thus, we must essentially perform two independent search  
processes, one for $\alpha$-models and one for $\beta$-models, and
we terminate these searches only when they have each reached  
the same degree of penetration as measured through their respective
numbers of attempts per model $d_\alpha/M_\alpha$.
The value of $N_\alpha/N_\beta$ obtained in this way should then be 
independent of the chosen reference value of $d_\alpha/M_\alpha$ 
for sufficiently large $d_\alpha/M_\alpha$. 
This method of extracting $N_\alpha/N_\beta$ is discussed more fully
in Ref.~\cite{Measure}, where the derivation and limitations 
of this method are outlined in detail.

Of course, in the process of randomly generating string models,
we cannot normally control whether a random new model is of the $\alpha$- or $\beta$-type.
Both will tend to be generated together, as part of the same random search.
Thus, our procedure requires that we completely {\it disregard}\/
the additional models of one type that might be generated in the
process of continuing to generate the required, additional models of the other type.
This is the critical implication of Eq.~(\ref{kgolden2}).
Rather than let our model-generating procedure continue for a certain
duration, with statistics gathered at the finish line as in Eq.~(\ref{naive}), we must instead
establish two separate finish lines for our search process, one for $\alpha$-models
and one for $\beta$-models.  Of course, these finish lines
are not completely arbitrary, and must be chosen such they correspond
to the same relative degree of penetration of the $\alpha$- and $\beta$-model
spaces.  
Indeed, these finish lines must be balanced so that they correspond to points
at which the same ratio of attempts per model  
has been reached.
However, these finish lines will not generally
coincide with each other, which requires that some data actually be disregarded
in order to extract meaningful statistical correlations.

As discussed in Ref.~\cite{Measure},
Eq.~(\ref{kgolden2}) will enable us to extract a value for the ratio $N_\alpha/N_\beta$ 
which is stable as a function of sample size
only when the biases within the $\alpha$-model space are the same as those within the
$\beta$-model space.
In such cases, we can refer to the physical characteristics 
$\alpha$ and $\beta$ as being in the same universality class.
However, for a given model-generation method (such as the free-fermionic construction
which we shall be employing in this paper), it turns out that many physical characteristics 
of interest  $\lbrace \alpha,\beta,...\rbrace$ have the property 
that they are in the same universality class.
In the rest of this paper, correlations for physical quantities will be quoted
only when the physical characteristics being compared are in the same universality class.
The above method is then used in order to extract these correlations.

\section{Supersymmetry on the heterotic landscape}
\setcounter{footnote}{0}
\label{General}

In this section, we begin our analysis of the structure of the heterotic string landscape.  
In so doing, we shall also provide an explicit example of the method described in Sect.~3. 
Our focus in this section is to determine the extent to which string models
with different levels of
unbroken supersymmetry ($\calN{=}0,1,2,4$) 
populate the tree-level four-dimensional heterotic landscape.
For $\calN{=}0$ models, we shall further distinguish between models which are
tachyon-free at tree level, and those which are tachyonic.
Note that these characteristics are all mutually exclusive and together
span the entire landscape of heterotic string models in
four dimensions.
Thus, our goal is to achieve nothing less than 
a partitioning of the full set of tree-level heterotic string models 
according to their degrees of supersymmetry.
(We stress that this analysis will be 
the only case in which unstable
tachyonic $\calN{=} 0$ string models will be considered in this paper.)
We will then proceed in Sect.~5 to examine questions related to correlations
between the numbers of unbroken supersymmetry generators and the corresponding gauge
groups.

The landscape of four-dimensional heterotic strings is a relatively large and complex structure.  
It may therefore be useful, as an initial step,
 to quickly recall the much smaller ``landscape'' of 
 {\it ten}\/-dimensional heterotic strings.
In ten dimensions, the maximal allowed supersymmetry is $\calN{=}1$, and thus
our tree-level ten-dimensional
landscape may be partitioned into only three categories:  $\calN{=}1$ models, $\calN{=}0$ tachyon-free models, 
and $\calN{=}0$ tachyonic models.
Note that since the $\calN{=}0$ tachyonic models are not even stable at tree level,
the tree-level ``landscape'' actually consists only of models in the first two categories.
However, for convenience, in this section we shall use the word ``landscape'' to describe the full set
of heterotic vacuum solutions regardless of stability.    

\begin{table}[htb]
\begin{center}
\begin{tabular}{||c||c|c||}
\hline
\hline
SUSY class & \% of 10D landscape  & \% of reduced 10D landscape\\
\hline 
$\calN {=} 0$ (tachyonic)&   66.7  &   62.5    \\
~~$\calN {=} 0$ (tachyon-free)~~&   11.1  &  12.5  \\
$\calN {=} 1$ &    22.2     &  25.0  \\
\hline
\hline
\end{tabular}
\end{center}
\caption{
  Classification of the ten-dimensional tree-level heterotic ``landscape'' as a function
  of the number of spacetime supersymmetries and the presence/absence of tachyons at
  tree level.
  As always, models are judged to be distinct based on their gauge groups
  and particle contents.
  The full ten-dimensional heterotic landscape consists of nine distinct string models,
  while the landscape of models accessible through our random search
  methods is reduced by one model.
  In either case, we see that two thirds of the 
  tachyon-free portion of the ten-dimensional landscape is supersymmetric.  
   Thus unbroken supersymmetry tends to dominate the ``landscape''
    consisting of ten-dimensional models which are stable at tree level. }
\label{10Dtable}
\end{table}

As is well known~\cite{KLTclassification}, the full set of 
 $D=10$ heterotic strings consists of nine distinct string models:  
two are supersymmetric [these are the $SO(32)$ and $E_8\times E_8$ models],
one is non-supersymmetric but tachyon-free [this is the $SO(16)\times SO(16)$ string
model~\cite{SOsixteen}], and six additional models are non-supersymmetric and tachyonic.
Expressed as proportions of a full ten-dimensional heterotic landscape,
we therefore find the results shown in the middle column of Table~\ref{10Dtable}. 
It is important to note, however, that not all of these models would be realizable
through the methods we shall be employing in this paper (involving a construction
in which all degrees of freedom are represented in terms of complex worldsheet fermions).
Indeed, one of the tachyonic non-supersymmetric models 
exhibits rank-reduction and thus would not be realizable in a random search
of the sort we shall be conducting.
Statistics for the corresponding ``reduced'' landscape of accessible models
are therefore listed along the third column of  Table~\ref{10Dtable}; 
these are the statistics which will form the basis for future comparisons.
Note that in either case, the tachyon-free portion of the ten-dimensional landscape is dominated 
by supersymmetric models.  
This suggests that breaking supersymmetry without introducing tachyons is relatively
difficult in ten dimensions.

\begin{figure}[htb]
\centerline{
   \epsfxsize 4.0 truein \epsfbox {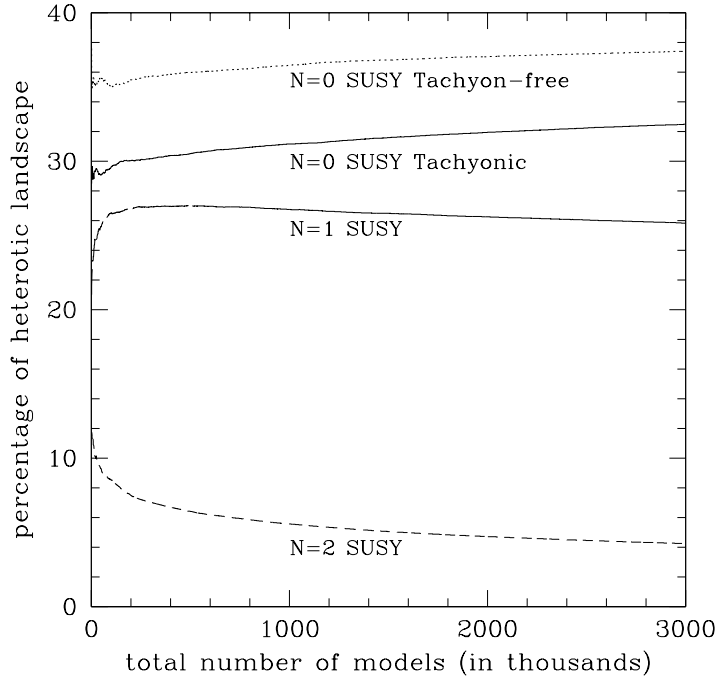}
    }
\caption{The numbers of distinct string models exhibiting different amounts of spacetime 
   supersymmetry, plotted as functions of the total number of distinct string models examined.  
   Models exhibiting $\calN{=}4$ supersymmetry are too few to appear on this figure.}  
\label{RunSUSYRun}
\end{figure}

Our goal is to understand how this picture changes after compactification to four dimensions.
Towards this end, one procedure might be to randomly generate a large set of 
string models, and see how many models one obtains of each type after a certain
fixed time as elapsed.  However, as discussed in Sect.~\ref{Analysis},
these percentages will generally
float or evolve as a function of the total number of models examined.
This behavior is shown in Fig.~\ref{RunSUSYRun}, and we see that
while the non-supersymmetric percentages seem to be floating towards
greater values, the supersymmetric percentages seem to be floating towards
lesser values.

As discussed in Sect.~\ref{Analysis}, it is easy to understand the reason
for this phenomenon.  Clearly, as we continue to generate models randomly,
an ever-increasing fraction of these models consists of models without supersymmetry.
This in turn suggests that at any given time, we have
already discovered a greater fraction of the space of supersymmetric models than 
non-supersymmetric models. 
This would explain why it becomes increasingly more difficult to randomly
generate new, distinct supersymmetric models as compared with non-supersymmetric
models, and why their relative percentages show the floating behavior
illustrated in Fig.~\ref{RunSUSYRun}. 

How then can we extract meaningful information?
As discussed in Sect.~\ref{Analysis}, the remedy involves keeping track 
of not only the total numbers of distinct models found in each supersymmetric class,
but also the total number of {\it attempts}\/ which yielded a model in each class,
even though such models were not necessarily new.
This information is shown in Table~\ref{secondtable}
for our total sample of $\gsim 10^{7}$ models.

\begin{table}[thb]
\begin{center}
\begin{tabular}{||c||r|r|r||}
\hline
\hline
SUSY class & $\#$ distinct models & $\#$ attempts & avg.\ attempts/model\\
\hline 
$\calN {=} 0$ (tachyonic)&   1$\,$279$\,$484   &    3$\,$810$\,$838  & 2.98 \\
$\calN {=} 0$ (tachyon-free)&   4$\,$946$\,$388  &    18$\,$000$\,$000 & 3.64\\
$\calN {=} 1$ &    3$\,$772$\,$679 &    24$\,$200$\,$097 & 6.41\\
$\calN {=} 2$ &   492$\,$790     &     13$\,$998$\,$843 &  28.41\\
$\calN {=} 4$ &  1106              &    6$\,$523$\,$277 & 5$\,$898.08\\
\hline
Total: & 10$\,$492$\,$447 & 66$\,$533$\,$055 & 6.34 \\
\hline
\hline
\end{tabular}
\end{center}
\caption{
This table expands on Table~\ref{firsttable} by including the numbers
of attempts to generate models in each class as well as the
corresponding average numbers of attempts per distinct model.
We also include information about the attempts which resulted in non-supersymmetric
models whose spectra are tachyonic at tree level.
It is apparent that the number of attempts per model increases rather
dramatically as the level of supersymmetry increases, indicating
that our heterotic string sample has penetrated further into the
spaces of models with greater numbers of supersymmetries than into those with fewer. }
\label{secondtable}
\end{table}

As we see from Table~\ref{secondtable},
the number of required attempts per model increases dramatically
with the level of supersymmetry.
This in turn implies, for example, that although we may have generated
many fewer distinct $\calN{=}4$ models  
than $\calN{=}1$ models,
the full space of $\calN{=}4$ models has already been penetrated
much more fully than the space of $\calN{=}1$ models. 
Thus, as we continue to generate more models, it should become relatively
easier to generate non-supersymmetric models than supersymmetric models.
If true, this would imply that the relative proportion of non-supersymmetric
models should increase as we continue to generate more models, while the
relative proportion of supersymmetric models should decrease.
This is, of course, exactly what we have already seen in Fig.~\ref{RunSUSYRun}. 

In order to extract final information concerning the relative sizes
of these spaces, 
the procedure outlined in Sect.~\ref{Analysis} instead requires that
we do something different, and
compare the numbers of distinct models generated in each  
class {\it at those points in our model-generating process
when their corresponding numbers of attempts per model are equal}\/.
It is only in this way that we can overcome the effects of 
floating correlations and extract stable relative percentages which do not
continue to evolve as functions of the total sample size.

For example, let us consider the relative numbers of $\calN{=}1$ and $\calN{=}2$
models.  Although we see from Table~\ref{secondtable} that 
our full sample of $\gsim 10^7$ models contains approximately
$7.66$ times as many $\calN{=}1$ models as $\calN{=}2$ models, this is not the relative
size of their corresponding model spaces because the 
$\calN{=}2$ space of models has already been explored more fully than the $\calN{=}1$
model space, with $6.41$ attempts per $\calN{=}1$ model compared with $28.41$
attempts per $\calN{=}2$ model.
However, at an earlier point in our search, we found that it took an average
of approximately $6.41$ attempts to generate a new, distinct $\calN{=}2$   
model:  this occurred when we had generated only approximately $90\,255$ models
with $\calN{=}2$ supersymmetry.  This suggests that the space
of $\calN{=}1$ models is actually $3772679/90255 \approx 41.8$ times as large as the
space of $\calN{=}2$ models.  

Moreover, we can verify that this ratio 
is actually stable as a function of
sample size.  For example, at an even earlier point in our search when
we had generated only $\approx 2.22\times 10^6$ $\calN{=}1$ models,
we found that an average of $3.64$ attempts were required to generate a new, distinct $\calN{=}1$ model.
However, this same average number of attempts per model occurred in our $\calN{=}2$ sample 
when we had generated only $\approx 53\,000$ $\calN{=}2$ models.
Thus, once again, the $\calN{=}1$ and $\calN{=}2$ model spaces appear to have
a size ratio of $\approx 41.8:1$. 

In this way, by comparing total numbers of models examined at equal 
values of attempts per model, 
we can extract the relative sizes of the spaces of models with differing
degrees of supersymmetry and verify that these results are stable as
functions of sample size (\ie, stable as functions of the chosen value
of attempts per model).
Our results are shown in Table~\ref{tthirdtable}.  ~As 
far as we can determine, the percentages quoted
in Table~\ref{tthirdtable}
represent the values to which the percentages in Fig.~\ref{RunSUSYRun}
would float if we could analyze  
what is essentially the full landscape.
However, 
short of examining the full landscape,
we see that there is no single point at which
these percentages would simultaneously appear
in any finite extrapolation of Fig~\ref{RunSUSYRun}.
Instead, it is only by comparing the numbers of models obtained at {\it different}\/ 
points in our analysis that the true ratios quoted in Table~\ref{tthirdtable} 
can be extracted.

\begin{table}[htb]
\begin{center}
\begin{tabular}{||c|c||}
\hline
\hline
SUSY class & \% of heterotic landscape\\
\hline
${\cal N}{=}0$ (tachyonic) & $32.1$\\
${\cal N}{=}0$ (tachyon-free) & $46.5$\\
${\cal N}{=}1$ & $20.9$\\
${\cal N}{=}2$  & $0.5$ \\
${\cal N}{=}4$  & $0.003$ \\
\hline
\hline
\end{tabular}
\end{center}
\caption{
   Classification of the four-dimensional tree-level heterotic landscape as a function of the 
   number of unbroken spacetime supersymmetries and the presence/absence of tachyons at
    tree level.  This table is thus the four-dimensional counterpart of 
    Table~\protect\ref{10Dtable}, which quoted analogous results for ten dimensions.
    Relative to the situation in ten dimensions, we see that compactification to four
    dimensions tends to {\it favor}\/ breaking all spacetime supersymmetries without introducing
    tachyons at tree level. }
\label{tthirdtable}
\end{table}

Table~\ref{tthirdtable} thus represents
our final partitioning of the tree-level four-dimensional landscape
according to the amount of supersymmetry exhibited. 
There are several rather striking facts which are evident from these results:
\begin{itemize}
\item{}  First, we see that nearly half of the heterotic landscape is 
             non-supersymmetric and yet tachyon-free.
\item{}  Second,  we see that the supersymmetric portion of the heterotic
           landscape appears to account for less than one-quarter of 
           the full four-dimensional heterotic landscape.
\item{}  Finally, models exhibiting extended ($\calN\geq 2$) supersymmetries are exceedingly
           rare, representing less than one percent of the full landscape.
\end{itemize}

Of course, we stress once again that these results hold only for the {\it tree-level}\/ landscape,
\ie, models which are stable at tree level only.
It is not clear whether these results would persist after full moduli stabilization. 
However, assuming that they do,
these results lead to a number of interesting conclusions.

The first conclusion is that the properties of the tachyon-free heterotic 
landscape as a whole are statistically 
dominated by the properties of string models which do {\it not}\/ have spacetime 
supersymmetry.  Indeed, the $\calN{=} 0$ string models account for over three-quarters 
of this portion of the heterotic 
string landscape.  The fact that the $\calN{=} 0$ string models dominate the tachyon-free portion of the 
landscape suggests that breaking supersymmetry without introducing tachyons 
is actually {\it favored}\/ over 
preserving supersymmetry for this portion of the landscape.  
Indeed, we expect this result to hold even after full moduli stabilization,
unless an unbroken supersymmetry is somehow restored by stabilization. 

The second conclusion which can be drawn from these results is that the 
supersymmetric portion of the landscape is almost completely comprised of
$\calN{=}  1$ string models.  Indeed, only  $2 \%$ of the supersymmetric 
portion of the heterotic landscape has more than $\calN{=} 1$ supersymmetry.  This suggests that 
the correlations present for the supersymmetric portion of the landscape
can be interpreted as 
the statistical correlations within the $\calN{=} 1 $ string models, 
 with the $\calN{=} 2$ correlations representing a correction at 
the level of $2\%$ and the $\calN{=} 4$ correlations representing a nearly negligible correction.

It is natural to ask what effects are responsible for this hierarchy.  
As was discussed in Sect.~\ref{Analysis}, 
two string models are considered distinct if any of their spacetime 
properties are found to be different.  Two models which have the same number
of unbroken spacetime supersymmetries must therefore differ 
in other features, such as their gauge groups and particle representations.
Thus, if there exist more models with one level of supersymmetry than another,
this must mean that there are more string-allowed configurations
of gauge groups and particle representations with one level of supersymmetry
than the other.
Indeed, given the results of Table~\ref{tthirdtable}, 
our expectation is that 
increasing the level of supersymmetry will have the effect of decreasing the number of 
distinct models with a given gauge group,
and possibly even the range of allowed gauge groups.
We shall test both of these expectations explicitly in Sect.~\ref{Specific}.

\section{Supersymmetry versus gauge groups}
\setcounter{footnote}{0}
\label{Specific}

Within the heterotic string, worldsheet self-consistency conditions
arising from the requirements of conformal anomaly cancellation, one-loop and multi-loop
modular invariance, physically sensible GSO projections, {\it etc}., 
impose many tight constraints on the allowed particle spectrum.
These constraints simultaneously affect not only the spacetime Lorentz structure of the theory
(such as is involved in spacetime supersymmetry), but also the 
internal gauge structure of the theory.
Thus, it is precisely within the context of string theory that we expect to 
find correlations between supersymmetries and gauge symmetries --- features which
would otherwise be uncorrelated in theories based on point particles.

In general, these correlations can lead to certain tensions in a given string
construction.  Models exhibiting large numbers of unbroken supersymmetries 
may be expected to have relatively rigid gauge structures, and vice versa.    
There are two specific types of correlations which we shall study.
First, we shall analyze how the degree of supersymmetry affects the range of
possible allowed gauge groups. 
For example, in extreme cases it may occur  
that certain gauge symmetries may not even be allowed for certain levels 
of spacetime supersymmetry.  
Second, even within the context of a fixed gauge group, we can expect the degree
of spacetime supersymmetry to affect the range of allowed particle representations which
can appear at the massless level. 
In other words, the number of distinct string models with a given fixed gauge group
may be highly sensitive to the degree of spacetime supersymmetry.

Some of these features are already on display in the ten-dimensional 
heterotic ``landscape''.  For example, no gauge group is shared
between those ten-dimensional models with supersymmetry and those without.
Moreover, in each case, there is only a single model with each
allowed gauge group.  Thus, in ten dimensions, the specification
of the level of supersymmetry (and/or the gauge group) is sufficient
to completely fix the corresponding particle spectrum.

Clearly, in four dimensions, things will be far more complex.
In particular, we shall study three correlations in this section: 
\begin{itemize}
\item First, 
   we shall focus on the number of allowed gauge groups 
    as a function of the degree of supersymmetry.
   We shall also study gauge-group multiplicities --- \ie, the probabilities 
   that there exist distinct string models with the same gauge group but different particle spectra.
       This will be the focus of Sect.~5.1.
\item Second, as a function of the degree of supersymmetry, we shall investigate
        ``shatter'' --- \ie, the degree to which our total 
         (rank-22) gauge group is ``shattered'' into distinct 
           irreducible factors, or equivalently the average
         rank of each irreducible gauge-group factor.   
                  This will be the focus of Sect.~5.2. 
\item Finally, as a function of the degree of supersymmetry,
        we shall study the probabilities of realizing specific (combinations of) gauge-group factors
        in a given string model.
           This will be the focus of Sect.~5.3.
\end{itemize}
As we shall see, these studies will find deep correlations which ultimately reflect the
string-theoretic tension between supersymmetry and the string consistency conditions.

\subsection{Numbers and multiplicities of unique gauge groups}
\setcounter{footnote}{0}

We begin by studying the total numbers of distinct gauge groups which can be realized
as a function of the number of unbroken supersymmetries in a given string model.

\begin{figure}[htb]
\centerline{
   \epsfxsize 4.0 truein \epsfbox {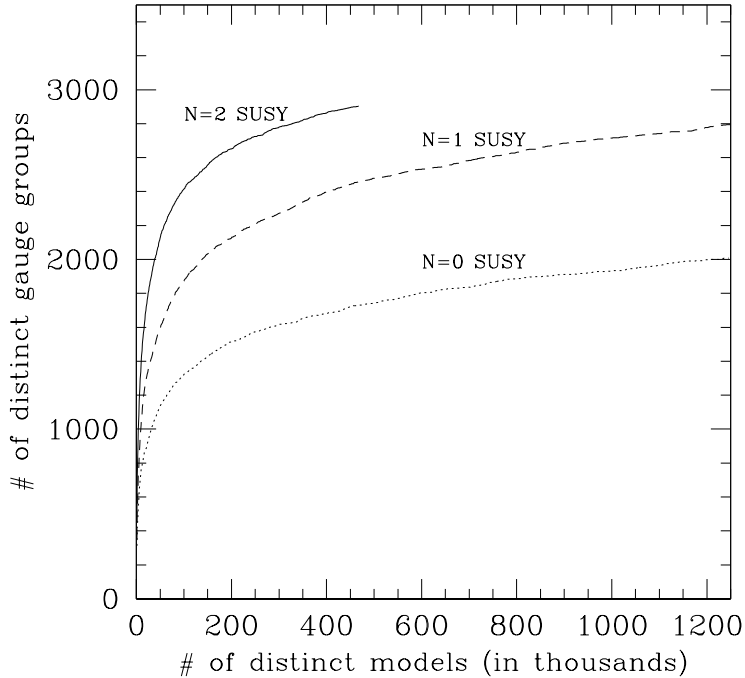}
    }
\caption{Numbers of distinct gauge groups obtained as functions
of the number of distinct string models generated.   Each curve
corresponds to models with a different number of unbroken spacetime
supersymmetries, with $\calN{=}0$ signifying models which are non-supersymmetric
but tachyon-free.  We see that for a fixed sample size, models 
with more unbroken supersymmetries tend to exhibit a larger number
of distinct gauge groups.  (Note that models with ${\cal N}{=}4$ supersymmetry
are too few to be shown in this plot.)}
\label{Bias} 
\end{figure}

To do this, one direct approach can might be to classify models according to their numbers of
unbroken spacetime supersymmetries, and tabulate the numbers of distinct gauge
groups which appear as functions of the total number of models in each class.
As we continue to generate more and more models, we then obtain the results 
shown in Fig.~\ref{Bias}.

It is evident from Fig.~\ref{Bias} that for a fixed sample size,
models
with more unbroken supersymmetries tend to exhibit larger numbers
of distinct gauge groups, or equivalently smaller numbers of model multiplicities
per gauge group.  For example, we see from Fig.~\ref{Bias}
that when each class of models has reached a sample size of $500\,000$ models,
the tachyon-free $\calN{=}0$ models have a greater multiplicity per
gauge group than $\calN{=}1$ models by an approximate factor $\approx 1.4$,
while the $\calN{=}2$ models have a smaller multiplicity per gauge
group than the $\calN{=}1$ models by an approximate factor $\approx 0.8$. 
However, it is easy to understand this
behavior.
As the level of supersymmetry increases, there are more constraints on the possible
particle spectra that can emerge for a given gauge group.
This in turn implies that there are likely to be fewer ways for two 
models with the same gauge group to be distinct,
which in turn  
implies that there is a greater chance that distinct models will be
forced to exhibit distinct gauge groups.
Thus, models exhibiting greater amounts of supersymmetry are likely, on average, 
to exhibit greater numbers of gauge groups 
amongst a fixed number of models.

Of course, as also evident from Fig.~\ref{Bias}, 
the multiplicity of distinct models per gauge group exhibits a strong, floating
dependence on the sample size.
Therefore, in order to extract a stable ratio of multiplicity ratios  --- one
which presumably represents the values of these ratios when extrapolated to the full landscape ---
we must employ the methods described in Sect.~\ref{Analysis}.
We then obtain the results shown in the middle column of Table~\ref{LandGauge}. 
Using these results in conjunction with the corresponding ratios of landscape 
magnitudes in Table~\ref{tthirdtable},
we can also calculate the relative numbers of distinct gauge groups realizable
within each SUSY class of models.
These results are shown in the final column of Table~\ref{LandGauge}. 
Note that in each case, these quantities
are quoted as ratios relative to their $\calN{=}1$ values;
this represents the most detailed information that can be extracted using the methods 
of Sect.~\ref{Analysis}.

\begin{table}[htb]
\begin{center}
\begin{tabular}{||c||c|c||}
\hline
\hline
 ~          & avg. multiplicity & \# of realizable \\
 SUSY class & per gauge group &  gauge groups \\
\hline
${\cal N}{=} 0$ (tachyon-free)  & 1.65  &   1.35     \\
${\cal N}{=} 1$                 & 1.00  &   1.00     \\
${\cal N}{=} 2$                 & 0.89  &   0.03      \\
\hline
\hline
\end{tabular}
\end{center}
\caption{The average relative multiplicities (distinct models per gauge group) 
and total numbers of realizable gauge groups, 
evaluated for heterotic string models with ${\cal N}=0,1,2$ unbroken
spacetime supersymmetries.  In each case, these quantities 
are normalized to their $\calN{=}1$ values.   }
\label{LandGauge}
\end{table}

We see from Table~\ref{LandGauge} that both the average multiplicities per gauge
group and the total numbers of realizable gauge groups
are monotonically decreasing functions of the number of unbroken supersymmetries.
While this is to be expected on the basis of the arguments described above,
we must realize that our class of $\calN{=}0$ models does not consist
of {\it all}\/ non-supersymmetric models, but merely those which are tachyon-free.
Thus, the requirement
of avoiding tachyons could have turned out to be more stringent than the requirement
of maintaining an unbroken supersymmetry,  at least as far as generating
a variety of gauge groups is concerned.
This is indeed what happens in the {\it ten}\/-dimensional 
landscape, where there are fewer realizable gauge groups for
non-supersymmetric tachyon-free models
than for models with $\calN{=}1$ supersymmetry.
However, the results in Table~\ref{LandGauge} indicate that the 
opposite is true in $D=4$.

Note that in Table~\ref{LandGauge},
we do not quote results for the $\calN{=}4$ portion of the
heterotic landscape because the absolute numbers
of models in this class are so small
that no stable numerical results can be extracted relative to the
other levels of supersymmetry.
However, it is worth noting that literally each $\calN{=}4$ model
in our sample has a unique gauge group, so the {\it absolute}\/ (rather than relative)
gauge-group multiplicity in the $\calN{=}4$ case is exactly $1.000$.  
This only reinforces our general observation that increased levels of supersymmetry
reduce the gauge-group multiplicity;  indeed, we now see that the case
of {\it maximal}\/ supersymmetry appears to result in the {\it minimal}\/ allowed
gauge-group multiplicity.
It is likely that this result can be proven analytically
for the $\calN{=}4$ landscape as a whole.

\subsection{Shatter/average rank}
\setcounter{footnote}{0}
\label{Shatter}

Having studied the numbers of different possible gauge groups,
we now turn our attention to the gauge groups themselves.
Once again, our goal is to study how these gauge groups depend on the
presence or absence of spacetime supersymmetry.
 
To begin the discussion,
our focus in this section will be on what we call ``shatter''~\cite{dienes}.
Recall that the heterotic string models we are considering 
all have gauge groups with total rank $22$.
This stretches from models with gauge group $SO(44)$ all the way down
to models with gauge groups of the form $U(1)^n \times SU(2)^{22 - n}$
with potentially all values of $n$ in the range $0\leq n\leq 22$.
Following Ref.~\cite{dienes},
we shall define the ``shatter'' for a given string model
as the number of
distinct irreducible gauge-group factors
into which its total rank-22 gauge group has been shattered.
Note that for this purpose, factors of $SO(4)\sim SU(2)\times SU(2)$ contribute two units
to shatter.  Since the total rank of the gauge group is fixed at 22 
for such models,
this means that shatter is also a measure of the average rank of the
individual group group factors, with $\langle \hbox{rank}\rangle= 22/\hbox{shatter}$.
Roughly speaking, shatter can also be taken as a measure of the degree of complexity
needed for the construction of a given string model, with increasingly smaller 
individual gauge-group factors tending to require increasingly many non-overlapping
sequences of orbifold twists and Wilson lines.

Given this definition of shatter, we may then calculate the distribution of shatter
across the landscape of heterotic strings.  We may calculate, for example, the
relative probabilities that models with certain levels of shatter emerge across
the landscape, and ask how these probability distributions vary with the amount
of spacetime supersymmetry present in the model.

\begin{figure}
\centerline{
   \epsfxsize 3.0 truein \epsfbox {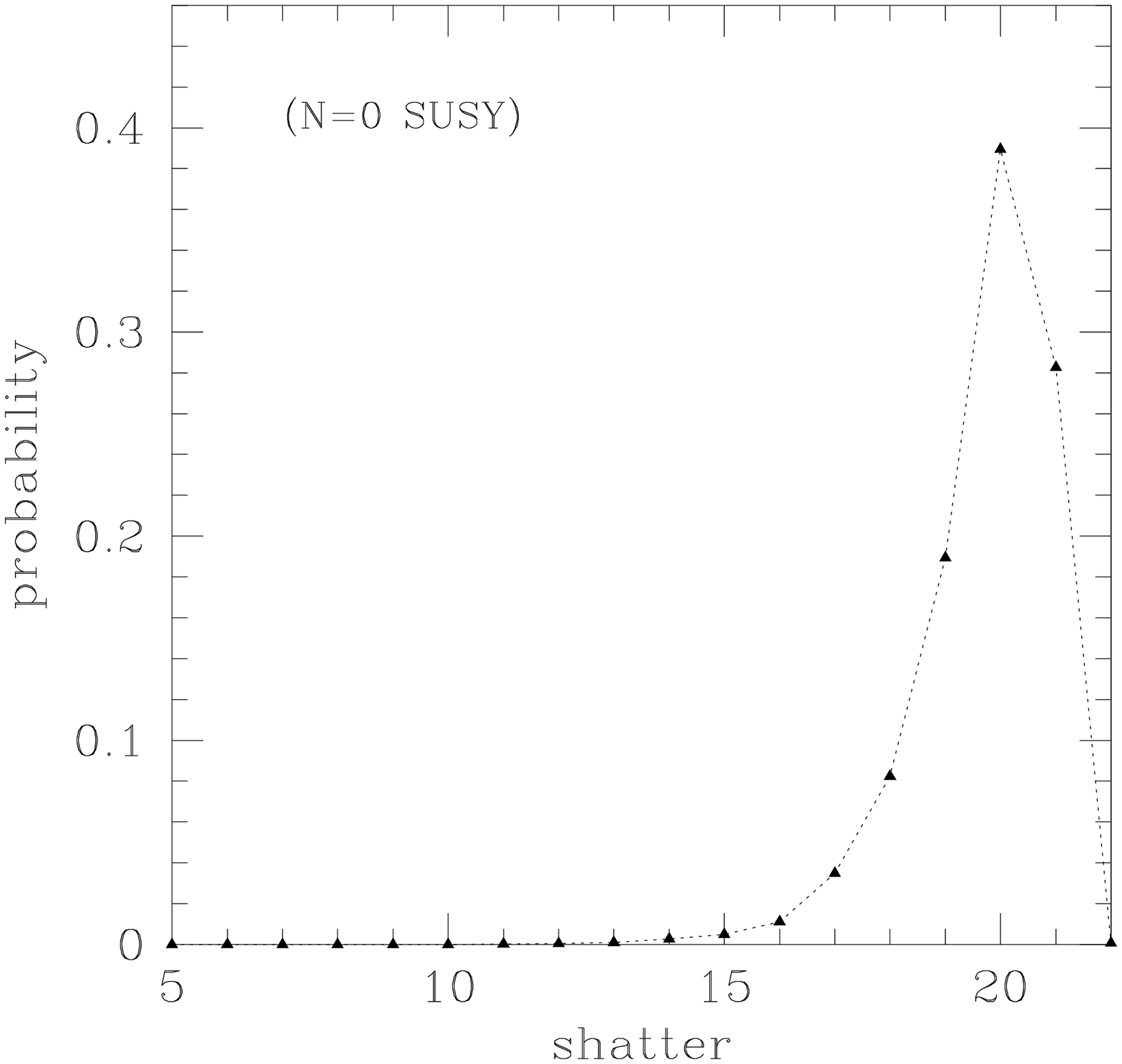}
    \hfill
   \epsfxsize 3.0 truein \epsfbox {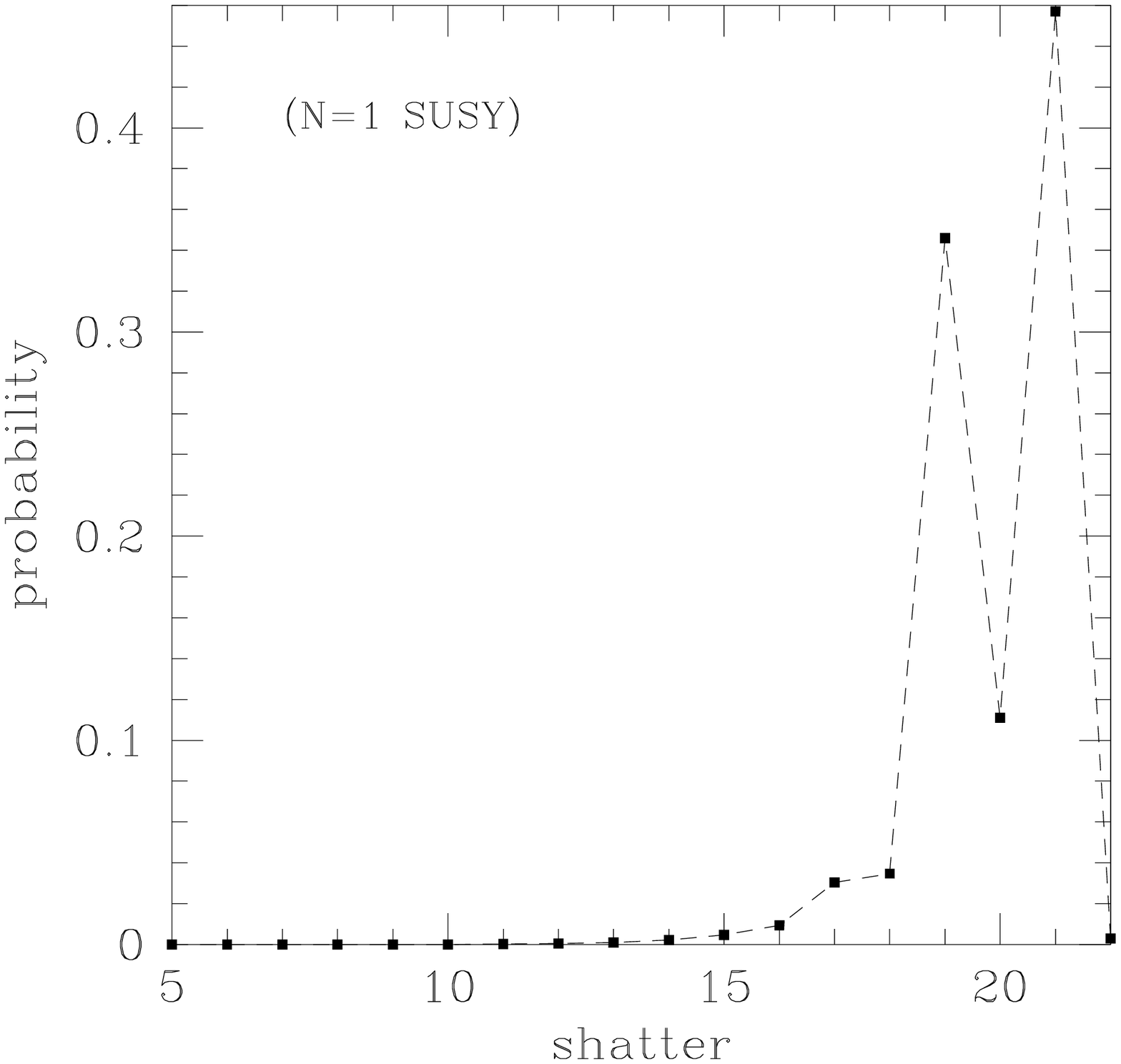}
   }
\centerline{
   \epsfxsize 3.0 truein \epsfbox {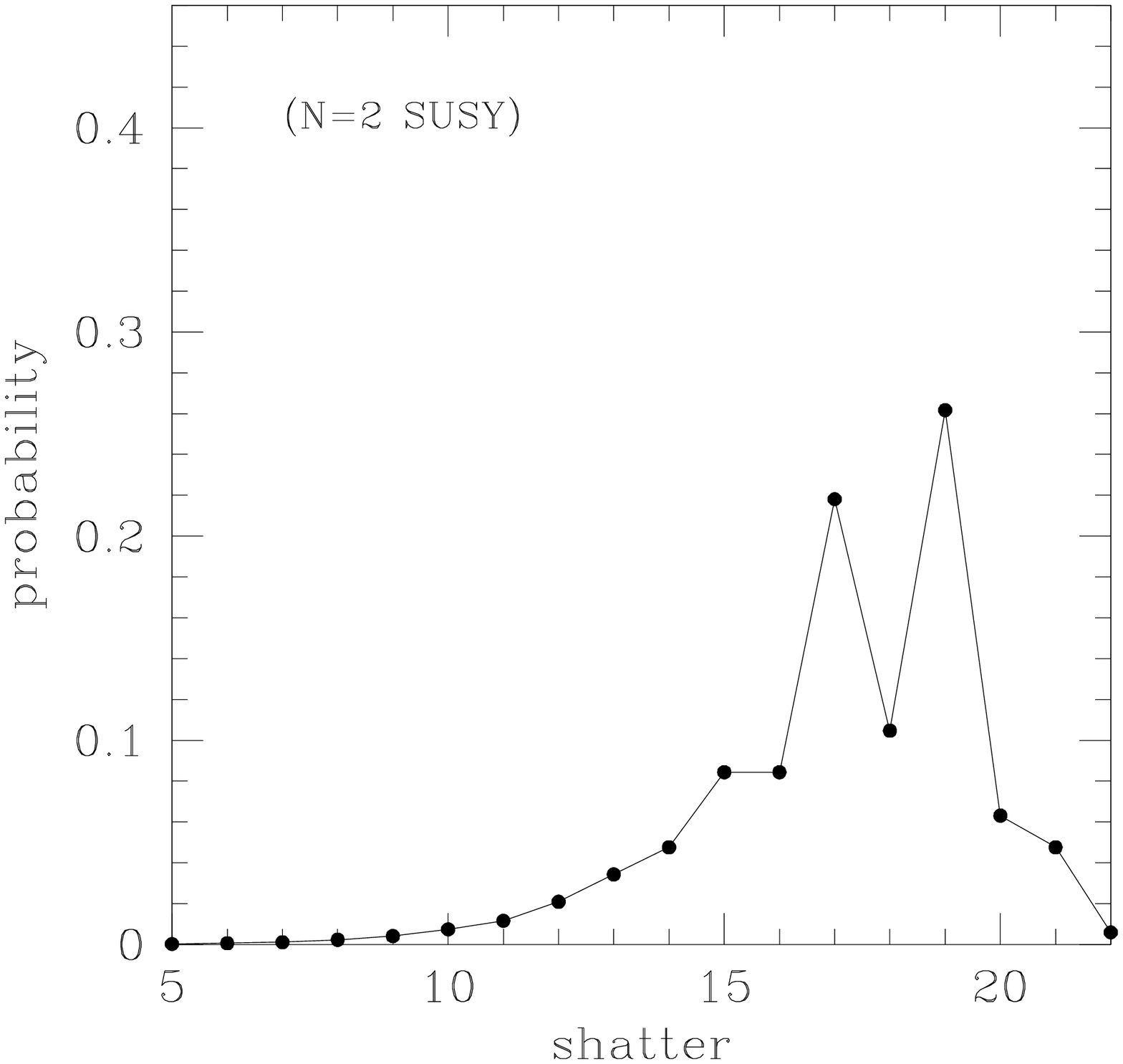}
    \hfill
   \epsfxsize 3.0 truein \epsfbox {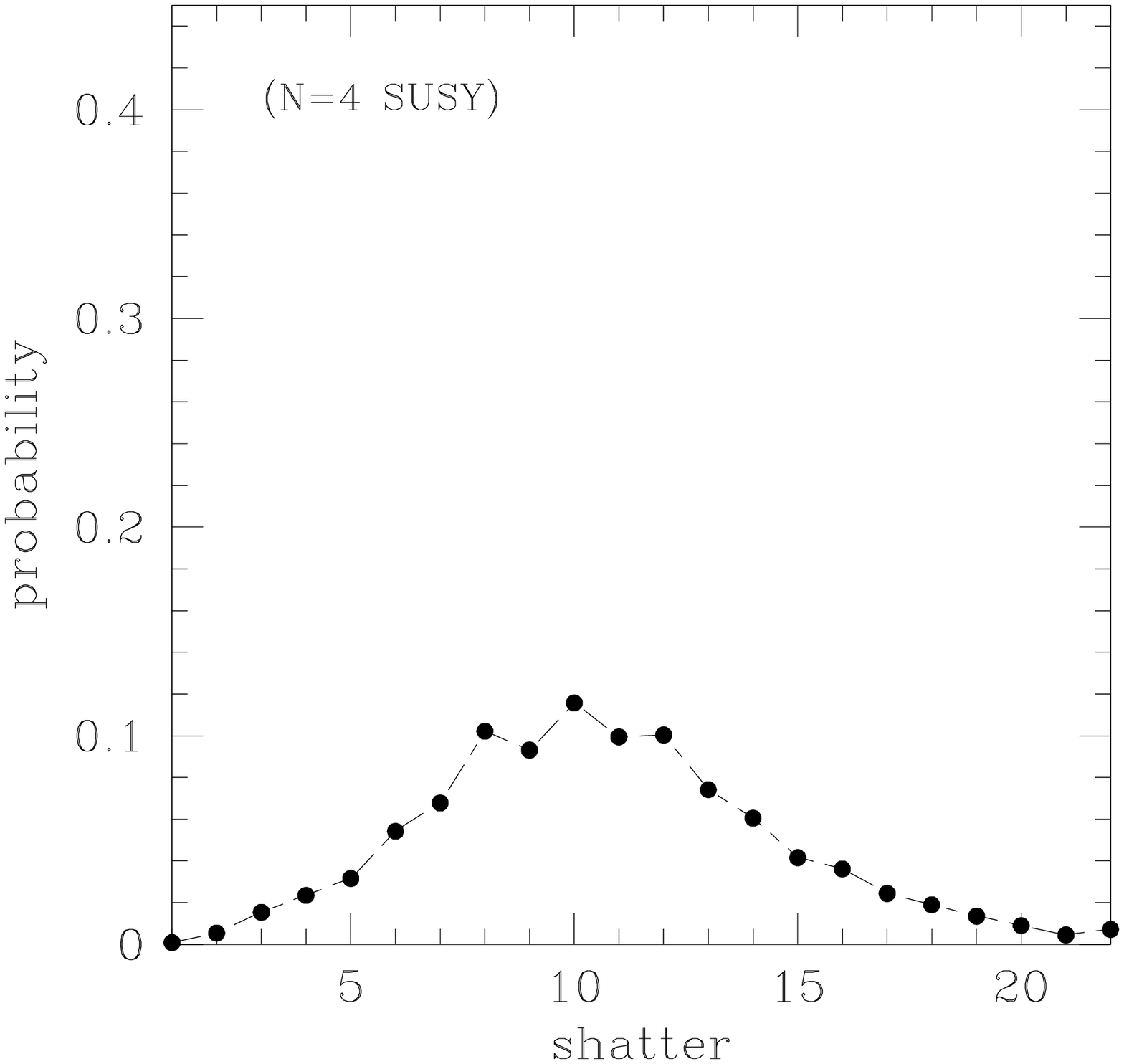}
    }
\caption{The absolute probabilities of obtaining distinct four-dimensional heterotic 
   string models with different numbers of unbroken supersymmetries,
   plotted as functions of the degree to which their gauge
   groups are ``shattered'' into separate irreducible factors.
    The total value of the points (the ``area under the curve'') in each case is 1.
    Here $\calN{=}0$ refers to models which are non-supersymmetric but tachyon-free.}
\label{Appleshatter} 
\end{figure}

Our results are shown in Fig.~\ref{Appleshatter}.
Once again, we stress that our raw data tends to evolve significantly as a function of
the sample size of models considered.  It is therefore necessary to employ the
techniques described in Sect.~3 in order to extract stable results which should
apply across the landscape as a whole.
In practice, this requires a difficult and time-consuming process
in which each of the data points shown in Figs.~\ref{Appleshatter}
for $\calN{=}{0,1,2}$ has individually been extracted through the limiting procedure described  
in Sect.~3.  Only then is an entire ``curve'' constructed for each
level of supersymmetry, as shown.

For the $\calN{=}4$ case, by contrast, our sample size is too small to permit stable
results to be extracted.  However, the fact that the attempts per model count in 
Table~\ref{secondtable} is so large for the $\calN{=}4$ models suggests that our $\calN{=}4$ sample
has already explored a significant fraction (and perhaps even most) of the corresponding landscape.
The $\calN{=}4$ curve in Fig.~\ref{Appleshatter} thus represents a direct tally of our
$\calN{=}4$ sample set.

As evident from Fig.~\ref{Appleshatter},
certain features of these plots are independent
of the level of spacetime supersymmetry.
These therefore represent general trends which hold across the entire tachyon-free 
heterotic string landscape.
For example, one general trend 
is a strong preference for models 
with relatively high degrees of shatter and correspondingly small 
average ranks for individual gauge-group factors --- models exhibiting
shatters near or in the teens clearly dominate.
On the other hand,
this preference for highly shattered gauge groups 
does {\it not}\/ appear to extend to the limit of completely shattered models with shatter${=}22$;
indeed, the set of models with only rank-one gauge-group factors seems to represent
a fairly negligible portion of the landscape regardless of the degree of supersymmetry.
This indicates that most models in this class have gauge groups which 
contain at least one factor of rank greater than one.\footnote{
       Of course, we stress that this conclusion applies only for
       models in the free-fermionic class.  In general, it is always
       possible to deform away from the free-fermionic limit by adjusting
       the internal radii of the worldsheet fields away from their free-fermionic
       values;  in such cases, we expect all gauge symmetries to be broken
       down to $U(1)^{22}$.  However, as noted earlier, the free-fermionic points 
      typically represent precisely those points at which additional 
     (non-Cartan) gauge-boson states become massless, thereby
    enhancing the gauge symmetries to become non-abelian.  Thus, as discussed more
    fully in Sect.~2, the 
    free-fermionic construction naturally leads to precisely the set of models 
    which are likely to be of direct phenomenological relevance.  }

Another universal trend implied by (though not explicitly shown in) Fig.~\ref{Appleshatter}
is that string models with shatters of less than four accrue
relatively little measurable amount of probability.  Even in the $\calN{=}4$ case,
these models are thus actually quite 
rare across the landscape as a whole.  In some sense, this too is to be expected,
since there are many more ways of breaking a large gauge symmetry through
orbifolds and non-trivial Wilson lines than of preserving it.
 
Despite these universal features, we see that spacetime supersymmetry
nevertheless does have a significant effect on the shapes of these curves. 
In this regard, there are two features to note.

First, we observe that as 
the degree of unbroken supersymmetry increases,
the {\it range}\/ of probable shatter values also tends to increase,
with probability shifting from models with high shatters to models with lower shatters.
This is especially noticeable when comparing the distribution of the $\calN{=}2$ and $\calN{=}4$ models with
those of the $\calN{=}0$ and $\calN{=}1$ models.
These results indicate that models exhibiting smaller amounts of shatter (\ie, models whose
gauge-group factors have larger individual average ranks) become somewhat
more probable as the level of supersymmetry increases.
Ultimately, this correlation between unbroken supersymmetry 
and unbroken gauge symmetry emerges since both have their underlying origins
in how our orbifold twists and Wilson lines are chosen.

Second, and perhaps more unexpectedly, we see that the degree of
supersymmetry also affects the overall profiles of these curves.
While the $\calN{=}0$ curve is relatively smooth, exhibiting a single peak
at shatter${=}20$, these curves begin to experience even/odd oscillations
as the degree of supersymmetry increases, with odd values of shatter significantly
favored over even values when supersymmetry is present.
The origins of this phenomenon are less apparent, and perhaps lie in the
modular invariance and anomaly cancellation
constraints which correlate the orders of the allowed twists 
leading to self-consistent string models. 
Interestingly, this even/odd behavior continues into the
$\calN{=}4$ case, although these oscillations are significantly less pronounced
and flip sign, with evens now dominating over odds.

One notable feature of the $\calN{=}4$ curve is its approximate reflection symmetry 
around shatter${=}10$.
It is unclear whether this is an exact symmetry which holds in situations
with maximal supersymmetry, or whether this is merely an accident.

\subsection{Specific gauge-group factors}
\setcounter{footnote}{0}
\label{Gauge}

Finally, we turn to an analysis of the probabilities of realizing individual gauge-group factors.
Just how likely is it, say, that a randomly chosen heterotic string model
will exhibit an $SU(3)$ factor in its gauge group, and how does this probability
correlate with the spacetime supersymmetry of the model?

\begin{figure}[htb]
\centerline{
   \epsfxsize 4.0 truein \epsfbox {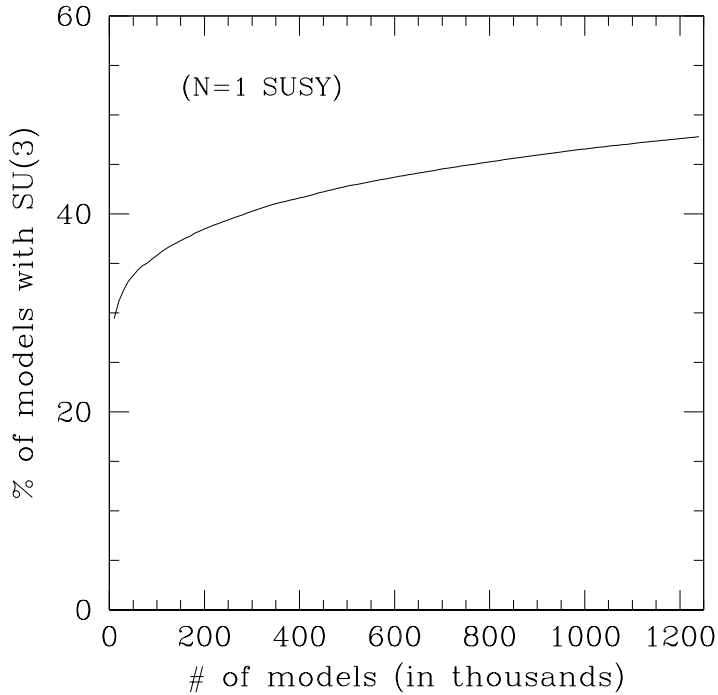}
    }
\caption{The percentage of distinct four-dimensional 
         ${\cal N}{=}1$ supersymmetric heterotic string models 
  exhibiting at least one $SU(3)$ gauge-group factor, plotted as 
  a function of the number of models examined for the first $1.25$~million 
   models.  We see that as we generate further models, 
   $SU(3)$ gauge-group factors become somewhat more ubiquitous --- 
    {\it i.e.}\/, the fraction of models with this property {\it floats}.    
   One must therefore account for this floating behavior
   using the methods described in Sect.~3 
   in order to extract meaningful information concerning
    the relative probabilities of specific gauge-group factors.}
\label{Bias2} 
\end{figure}

Just as with previous questions, addressing this issue requires a detailed analysis along
the lines discussed in Sect.~3.  This is because the probabilities of realizing different gauge-group
factors also float quite strongly as a function of sample size.
As dramatic illustration of this fact, let us restrict our attention to
models with $\calN{=}1$ spacetime supersymmetry
and calculate the probability that a given model will exhibit an $SU(3)$ gauge-group factor
as a function of the number of models we have examined.
We then obtain the result shown in Fig.~\ref{Bias2},
and it is clear that the percentage of models 
with $SU(3)$ gauge-group factors {\it floats}\/ rather significantly as a function 
of the sample size.  
Indeed, on the basis of this information alone, it would be quite impossible
to determine the final value to which this curve might float.
Just as with previous examples,
this floating behavior ultimately occurs
because models with $SU(3)$ gauge-group factors are relatively
difficult to generate using the construction methods we are employing;
thus, they tend to emerge in increasing numbers only after other models 
are exhausted.  As discussed more fully in Ref.~\cite{Measure},
this does not imply that
there are fewer of these models or that our construction method cannot ultimately
reach them --- all we can conclude is that they are less likely to be generated
in a random search than other models, and thus they tend to emerge
only later in the search process.
Indeed, as we shall shortly see, models with $SU(3)$ gauge-group factors 
actually tend to dominate the landscape.

Therefore, in order to extract meaningful results, we again employ
the methods discussed in Sect.~3.
We then obtain the final percentages quoted in Table~\ref{ApGroup}.  
We observe, in particular,  
that the probability of models 
with $\calN{=}1$ supersymmetry
exhibiting at least one $SU(3)$ gauge-group factor
has actually risen all the way to 98\%. 
The fact that this probability has floated from
nearly $55 \%$ to $98 \%$
only reinforces the importance of the analysis method presented 
in Sect.~\ref{Analysis}, and illustrates the need
to properly account for floating correlations
when quoting statistical results for such studies.

\begin{table}[htb]
\begin{center}
\begin{tabular}{||c||r|r|r|r||}
\hline
\hline
gauge group& ${\cal N}{=}0$&~${\cal N}{=}1$~&~${\cal N}{=}2$~&~${\cal N}{=}4$~\\
\hline
\hline
$U_1$ &99.9 & 94.5 & 68.4 &89.6\\
\hline
$SU_2$ & 62.46 & 97.4 & 64.3 & 60.9\\
\hline
$SU_3$ & 99.3 & 98.0  & 93.0  & 45.1\\
\hline
$SU_4$& 14.46 & 30.0 & 39.0  &53.5\\
\hline
$SU_5$&16.78 & 43.5 & 66.3 & 33.8\\
\hline
$SU_{>5}$ &0.185 &1.7 & 10.6 &73.0\\
\hline
$SO_8$& 0.482 & 1.6  & 6.2  & 21.1\\
\hline
$SO_{10}$&0.084  & 0.2  & 1.6  & 18.7\\
\hline
$SO_{>10}$ & ~$0.005$ &0.038  & 0.77 & 7.5\\
\hline
$E_{6,7,8} $& ~$0.0003$ & 0.03 & 0.16 & 11.5\\
\hline
\hline
\end{tabular}
\end{center}
\caption{Percentages of heterotic string models exhibiting specific 
          gauge-group factors as functions of their spacetime supersymmetry.  
            Here $SU_{>5}$ and $SO_{>10}$ collectively indicate gauge groups 
            $SU(n)$ and $SO(2n)$ for any $n>5$, while
             ${\cal N}$ refers to the number of unbroken supersymmetries at the string scale.
            Note that the $\calN{=}0$ models are all tachyon-free.}
\label{ApGroup}
\end{table}

As we see from Table~\ref{ApGroup}, 
supersymmetry can have quite sizable effects upon the probability 
of realizing specific groups.  However, there are some general trends that hold
for the full heterotic landscape.  These 
trends include:
\begin{itemize}
\item{}  A preference for $SU(n+1)$ over $SO(2n)$ groups for each rank $n$.  
         Even though these two groups have the same rank, it seems that
         $SU$ groups are more common than the $SO$ groups for all levels of 
         supersymmetry.
\item{}   Groups with smaller rank are much more common 
            than groups with larger rank.  Once again, this also appears to hold
         for all levels of supersymmetry. 
\item{}   Finally, the gauge-group factors comprising Standard-Model gauge group 
           $G_{\rm SM}\equiv SU_3\times SU_2\times U_1$
          are particularly common, much more so than those of any of its grand-unified extensions.
\end{itemize}

As we found in Sect.~4,
the $\calN{=} 0$ string models dominate the tachyon-free portion of the heterotic 
landscape.  Similarly,  the $\calN{=} 1$ string models are the dominant part of the supersymmetric portion of 
the landscape.  Nevertheless, it is interesting to examine the gauge-group probabilities across both of these 
portions of the landscape.  These probabilities are easy to calculate by combining the results 
in Tables~\ref{tthirdtable} and \ref{ApGroup},
leading to the results shown in Table~\ref{LandGroup}.

\begin{table}
\begin{center}
\begin{tabular}{||c||c|c||}
\hline
\hline
gauge& entire & SUSY\\
group& landscape   & subset \\
\hline
\hline
$U_1$ &98.00 & 93.89 \\
\hline
$SU_2$ & 73.22& 96.62 \\
\hline
$SU_3$ & 98.85 & 97.88  \\
\hline
$SU_4$& 19.42 & 30.21 \\
\hline
$SU_5$& 25.37 & 44.03\\
\hline
$SU_{>5}$ &0.73& 1.92 \\
\hline
$SO_8$& 0.87 & 1.71  \\
\hline
$SO_{10}$&0.13 & 0.23 \\
\hline
$SO_{>10}$ & 0.02 & 0.06  \\
\hline
$E_{6,7,8} $& 0.01& 0.03\\
\hline
\hline
\end{tabular}
\end{center}
\caption{Percentage of heterotic string models exhibiting specific
          gauge-group factors, quoted across the entire landscape of tachyon-free models 
              (both supersymmetric
          and non-supersymmetric) as well as
          across only that subset of models with at least $\calN{\geq}1$ spacetime supersymmetry.
         These results are derived from those of Table~\protect\ref{ApGroup}
         using the landscape weightings in Table~\protect\ref{tthirdtable}.}
\label{LandGroup}
\end{table}

Several features are immediately apparent from Table~\ref{LandGroup}.
First, gauge groups with larger ranks appear to be favored more strongly
          across the supersymmetric subset of the landscape 
      than across the tachyon-free landscape as a whole.
        Since each of our heterotic string models in this class has a gauge group of fixed 
         total rank, this preference for higher-rank gauge groups 
         necessarily comes at the price of sacrificing smaller-rank gauge groups.  
      Indeed, it often happens that 
this preference for larger-rank gauge groups actually precludes
the appearance of any small-rank gauge groups whatsoever.
Interestingly, the supersymmetric portion of the landscape seems to sacrifice $U(1)$ 
primarily and $SU(3)$ to a lesser extent.  This is in contrast to $SU(2)$, which is actually more strongly {\it 
favored}\/ in the supersymmetric portion of the landscape than in the general tachyon-free landscape
as a whole.

Second, the level of supersymmetry also appears to affect
        the probability distributions across the different possible gauge-group factors.  
        The supersymmetric portion of the landscape has a much greater representation of the large rank 
groups.  This suggests that the constraints placed on the string spectrum in order to preserve spacetime 
supersymmetry also have the effect of favoring larger gauge symmetries, a fact already noted 
in Sect.~5.2.~  In other words, there tends to be a decrease in the gauge-group multiplicity 
          for highly shattered gauge groups which consist of only very small gauge-group factors, 
          and thus the larger-rank gauge 
       groups make up a larger proportion of the whole landscape.  
        Indeed, this effect is particularly acute for that subset of the landscape
          exhibiting maximal $\calN{=}4$ supersymmetry, where the
           larger-rank $SU$ gauge groups are particularly well represented.

\section{Discussion}
\setcounter{footnote}{0}
\label{Conclusions}

In this paper, we have examined both the prevalence of spacetime supersymmetry across the heterotic
string landscape and the statistical correlations between the appearance of spacetime supersymmetry
and the gauge structure of the corresponding string models.
Somewhat surprisingly, we found that 
nearly half of the heterotic landscape is
non-supersymmetric and yet tachyon-free at tree level;  indeed, less than a quarter
of the tree-level heterotic landscape exhibits any supersymmetry at all at the string scale.
Moreover, we found that 
the degree of spacetime supersymmetry
is strongly correlated with the probabilities of realizing
certain gauge groups, with unbroken supersymmetry at the string scale
tending to favor gauge-group factors with larger rank.

There are several extensions to these results which are currently under investigation.
For example, we would like to understand how the presence of supersymmetry affects the
statistical appearance of the entire composite Standard-Model gauge group 
$G_{\rm SM}\equiv SU_3\times SU_2\times U_1$, and not merely  
the appearance of its individual factors.  We would also like to understand how the presence or
absence of supersymmetry affects other features which are equally important
for the overall architecture of the Standard Model:  these include the appearance
of three chiral generations of quarks and leptons, along with a potentially correct set of gauge  
couplings and Yukawa couplings.  This work has already been completed, and will
be reported shortly~\cite{toappear}.

Despite this progress, such studies have a number of intrinsic limitations which 
must continually be borne in mind.
A number of these have been emphasized by us in recent articles (see, \eg, the concluding sections
of Refs.~\cite{dienes,Measure}) and will not be repeated here.
However, other limitations are particularly relevant for the results we have quoted here
and thus deserve emphasis.

First, 
we must continually bear in mind that our study has been limited to models in which 
rank-cutting is absent.
Thus, all of the four-dimensional heterotic string models we have examined exhibit
a fixed maximal rank${=}22$.
This has the potential to skew the statistics of the different 
gauge-group factors.  For example, it is possible that 
gauge-group factors with smaller ranks 
might be over-represented in this sample simply because the appearance of such groups
is often the only way in which a given model can precisely saturate the
total rank bound.
By contrast, for models which can exhibit rank-cutting, this saturation would not
be needed and it is therefore possible that lower-rank groups are consequently
less abundant.

A second limitation of this study stems from the nature of performing random search studies
in general.  
In Sect.~\ref{Analysis}, we summarized several methods by which the problematic issue of floating
correlations can be transcended, and this paper has provided several examples of not only the need for such
methods but also of the means by which they are implemented.  As more fully discussed in Ref.~\cite{Measure},
such problems are going to arise --- and such methods are going to be necessary ---  whenever we 
attempt to extract statistical correlations from a large data set to which
our computational access is necessarily limited. 
However, despite the apparent success of such methods, it is always a logical possibility that 
there exists a huge pool of string models with non-standard physical characteristics
remaining just beyond our computational power to observe.  The existence of such a pool
of models would completely change the nature of our statistical results, to an extent
which is essentially unbounded, yet we may miss this completely because of limited
computational power.  Although this becomes increasingly unlikely as our search through the
landscape becomes larger and increasingly sophisticated,  
this nevertheless always remains a logical possibility which cannot be discounted.

But finally, perhaps the most serious limitation of our study is the fact that we are
analyzing the statistical properties of string models
which are not necessarily stable beyond tree level.  
Indeed, since none of our non-supersymmetric string models has a vanishing 
one-loop cosmological constant, these models in particular 
necessarily have non-zero dilaton tadpoles
at one-loop order and thus become unstable.
Even our supersymmetric models have flat directions which have not been lifted.
Thus, as we have stressed throughout this paper, the ``landscape'' we have examined in this
paper is at best a tree-level one.
Despite this fact, however, it is important to realize that 
these models do represent self-consistent 
string solutions at tree level.
Specifically, these models satisfy all of the 
constraints needed for
worldsheet conformal/superconformal invariance,
modular-invariant one-loop and multi-loop amplitudes,
proper spacetime spin-statistics relations,
and physically self-consistent layers of sequential GSO projections and orbifold twists.
Indeed,
since no completely stable perturbative heterotic strings have
yet been constructed, this sort of analysis is currently the state of the art
for large-scale statistical studies of this type.
This mirrors the situation on the Type~I side, where state-of-the-art
statistical analyses
have also focused on models which are only stable at tree level.

Nevertheless, we are then left with the single over-arching question:
to what extent can we believe that the results we have found for this
``tree-level'' landscape actually apply to the true landscape 
that would emerge after all moduli are stabilized?
The answer to this question clearly depends on the extent to which
the statistical correlations
we have uncovered here are likely to hold even after vacuum stabilization.

 {\it A priori}\/, this is completely unknown.
However, one surprising result of this paper is the observation
that the string self-consistency requirements themselves --- even merely
at tree-level --- do {\it not}\/ preferentially give rise to supersymmetric solutions
at the string scale.
Indeed, as we discussed in Sect.~4,
less than a quarter
of the tree-level heterotic landscape appears to exhibit 
any supersymmetry at all at the string scale.
Thus, breaking supersymmetry without introducing tachyons is actually 
statistically {\it favored}\/ 
over preserving supersymmetry,
even at the string scale and even when the requirements of avoiding tachyons are implemented.
Observations such as these tend to shift the burden of proof onto
the SUSY enthusiasts,
and perhaps reframe the question 
to one in which we might ask whether 
an unbroken supersymmetry is somehow {\it restored}\/ by modulus stabilization.
This seems unlikely, especially since most modern methods of modulus stabilization
rely on breaking rather than introducing supersymmetry.
In either case, however, this shows how the results of such studies --- even though
limited to only the tree-level landscape ---
can have the power to dramatically reframe the relevant questions.
Indeed, once the technology for building heterotic string models
develops further and truly stable vacua can be statistically analyzed in
large quantities,
it will be interesting to compare the statistical properties of
those vacua with these
in order to ascertain the degree to which vacuum stabilization
might affect these other phenomenological properties.

Thus, it is our belief that such statistical landscape studies of this sort
have their place, particularly when the results of such studies are 
interpreted correctly and in the proper context.
As such, we hope that this study of the perturbative
heterotic landscape may represent one small step in this direction.

\section*{Acknowledgments}
\setcounter{footnote}{0}

The work of KRD, ML, and VW 
is supported in part by the U.S. National Science Foundation
under Grant PHY/0301998,
by the U.S. Department of Energy under
Grant~DE-FG02-04ER-41298, and by a Research Innovation Award from
Research Corporation.

\bigskip
\vfill\eject
\bibliographystyle{unsrt}

\end{document}